\documentclass[usenatbib,onecolumn]{mn2e}
\usepackage[dvips]{graphicx}
\usepackage{colortbl}
\usepackage{amssymb,txfonts}
\usepackage{natbib}
\usepackage{bm}

\newcommand{\apj}{{ApJ}}
\newcommand{\apjs}{{ApJS}}
\newcommand{\mnras}{{MNRAS}}

\def\ee{\end{equation}}
\def\be{\begin{equation}}

\title[Global Neutrino Heating in Hyperaccretion Flows]{Global Neutrino Heating in Hyperaccretion Flows}

\author[S. Luo and F. Yuan]{Shu Luo$^{1}$\thanks{email: luoshu@xmu.edu.cn} and
Feng Yuan$^{2,1}$\thanks{email: fyuan@shao.ac.cn}\\
$^1$Department of Astronomy and Institute of Theoretical Physics and Astrophysics, Xiamen University, Xiamen, Fujian 361005, China\\
$^2$Key Laboratory for Research in Galaxies and Cosmology, Shanghai Astronomical Observatory, \\
Chinese Academy of Sciences,80 Nandan Road, Shanghai 200030, China\\}

\pagerange{\pageref{firstpage}--\pageref{lastpage}}
\pubyear{2013}

\begin{document}
\maketitle
\label{firstpage}
\begin{abstract}

The neutrino-dominated accretion flow (NDAF) with accretion rates $\dot{M}  = 0.01 - 10 ~ M^{}_{\sun} {\rm s}^{-1}_{}$ is a plausible candidate for the central engine of gamma-ray bursts (GRBs). This hyperaccretion disk is optically thin to neutrinos in the radial direction, therefore the neutrinos produced at one radius can travel for a long distance in the disk. Those neutrinos can thus be absorbed with certain probability by the disk matter at the other radius and  heat the disk there. The effect of this  ``global neutrino heating'' has been ignored in previous works and is the focus of this paper. We find that around the ``ignition'' radius $r_{\rm ign}$, the global neutrino heating rate could be comparable to or even larger than the local viscous heating rate thus must be an important process. Two possible consequences are in order if the ``global neutrino heating'' is taken into account: i) the temperature of the disk is slightly raised and the ``ignition'' radius $r_{\rm ign}$ slightly shifts to a larger radius, both lead to the increasing of the total neutrino flux; ii) what is more interesting is that, the temperature of the ADAF just beyond $r_{\rm ign}$ may be raised above the virial temperature thus the accretion will be suppressed. In this case, the activity of the black hole is expected to oscillate between an active and inactive phases. The timescale of the active phases is estimated to be $\sim 1$ second. If the timescale of  the inactive phase is comparable to or less than this value, this intermittent activity may explain the slow variability component of the GRBs. Self-consistent global calculations of NDAFs with the ``global neutrino heating'' included are required in the future to more precisely evaluate this effect.

\end{abstract}


\section{Introduction}

The accretion disks around black holes with rather high accretion rates $\dot{M}  = 0.01 - 10 ~ M^{}_{\sun} {\rm s}^{-1}_{}$ are plausible candidates for the central engine of gamma-ray bursts (GRBs) \citep{1992ApJ...395L..83N, 1999ApJ...518..356P, 2001ApJ...557..949N}. A large amount of neutrinos can be produced in such hyperaccretion flows. They will then collide, annihilate, and produce a fireball \citep{2000ApJ...530..292M}, which is one popular model for GRBs. In addition to the fireball model, another  type of models for the central engine of GRBs is magnetically dominated ones. Recent {\it Fermi} observations seem to suggest that the latter is more favored \citep{2009ApJ...700L..65Z}. The leading magnetic model is the Blandford-Znajek (BZ) mechanism \citep{1977MNRAS.179..433B}, which
requires a large-scale open magnetic field connecting the black hole
and an external astrophysical load. This model also
tends to generate a continuous jet. It is shown in this case that the energy dissipation via magnetized shocks in the jet is not efficient enough to explain observations \citep{2011MNRAS.416.2193N}. The magnetic turbulent reconnection is proposed instead \citep{2011ApJ...726...90Z}. Stimulated by the observations of episodic jets, \citet{2012ApJ...757...56Y} recently proposed a magnetic episodic jet model for GRBs. Different from the BZ mechanism, in this model, the jet is intrinsically episodic, i.e., in the form of discrete magnetized blobs. Reconnection can be easily triggered when they collide and GRBs are then naturally produced.

In this paper, we focus on the fireball model. The calculations by \cite{2007ApJ...657..383C} have shown that this hyperaccretion disk has an ``ignition'' radius $r^{}_{\rm ign}$ within which radius the neutrino production become efficient and the disk is neutrino-cooling dominated; while outside which radius the accretion flow is advection dominated. It has been found that except in the very inner part of the disk (i.e. $r < 10 \; r^{}_{s}$, where $r^{}_{s} = 2 G M / c^{2}_{}$ is the Schwarzschild radius of the black hole), the disk is optically thin to neutrinos in the radial direction, therefore the neutrinos produced at one radius can travel for a long distance in the disk before they are absorbed or scattered. These neutrinos can thus be absorbed with certain probability by the disk matter at the other radius and heat the disk there, as we will show below in this paper. We call this effect ``global neutrino heating''. This process has been ignored in previous works. In this paper we focus on this ``global neutrino heating'' in the neutrino cooling accretion flow, we find that around the ``ignition'' radius $r^{}_{\rm ign}$, the global neutrino heating rate could be comparable to or even stronger than the local viscous heating rate thus is non-negligible. We note that similar ``global Compton scattering'' effect of photons has been studied in the context of optically thin advection-dominated accretion flow \citep{2009ApJ...691...98Y}.

In section 2, we briefly review the accretion disk model that we are considering. Then in section 3 we show how to calculate the global neutrino heating rate and the  neutrino radiation pressure. Section 4 is devoted to the numerical results. A brief summary and discussion are given in section 5.

\section{The structure of accretion flow and the production of neutrinos}

Consider a stellar-size Schwarzschild black hole (throughout this paper we set the black hole mass $M = 3 ~ M^{}_{\sun}$ and the spin $a = 0$) surrounded by a hyper-accreting accretion flow with a typical accretion rate $\dot{M}  = 0.1 - 1.0 ~ M^{}_{\sun} {\rm s}^{-1}_{}$. Within the ignition radius $r^{}_{\rm ign}$, the neutrino cooling is important and the disk can be regarded as a neutrino-dominated accretion flow (NDAF). The disk properties in this region can be approximately described by \citep{1999ApJ...518..356P, 2006ApJ...643L..87G, 2007ApJ...657..383C, 2007ApJ...661.1025L, 2007ApJ...662.1156K}
\begin{eqnarray}
T & = & 1.06 \times 10^{11}_{} \; \alpha^{0.2}_{} \; M^{-0.2}_{} \; R^{-0.3}_{} ~ {\rm K} \; , \\
H & = & 4.33 \times 10^{4}_{} \; \alpha^{0.1}_{} \; M^{0.9}_{} \; R^{1.35}_{} ~ {\rm cm} \; , \\
\rho & = & 0.205 \times 10^{14}_{} \; \dot{M} \; \alpha^{-1.3}_{} \; M^{-1.7}_{} \; R^{-2.55}_{} ~ {\rm g} \cdot {\rm cm}^{-3}_{} \; , \\
\label{density}
\Omega & = & \Omega^{}_{K} \; = \; 0.707 \times 10^{5}_{} \; M^{-1}_{} \; R^{-1.5}_{} ~ {\rm s}^{-1}_{} \; , \\
v & = & 6.43 \times 10^{8}_{} \; \alpha^{1.2}_{} \; M^{-0.2}_{} \; R^{0.2}_{} ~ {\rm cm} \cdot {\rm s}^{-1}_{}.
\end{eqnarray}
At the outer disk, $r > r^{}_{\rm ign}$, when the temperature falls down and the mean electron energy becomes lower than $(m^{}_{n} - m^{}_{p}) c^{2}_{}$, the neutrino emission then rapidly switches off and the neutrino cooling becomes inefficient. At this region, the disk thickness is reduced and the cooling is dominated by advection. The properties of the advection-dominated accretion flow (ADAF) can be approximately estimated by \citep{1999ApJ...516..420W,2012ApJ...757...56Y}
\begin{eqnarray}
T & = & 3.47 \times 10^{11}_{} \; \dot{M}^{0.25}_{} \; \alpha^{-0.25}_{} \; M^{-0.5}_{} \; R^{-0.625}_{} ~ {\rm K} \; , \\
H & \approx & r / \sqrt{5} \; = \; 1.32 \times 10^{5}_{} \; M^{}_{} \; R^{}_{} ~ {\rm cm} \; , \\
\rho & = & 5.37 \times 10^{11}_{} \; \dot{M} \; \alpha^{-1}_{} \; M^{-2}_{} \; R^{-1.5}_{} ~ {\rm g} \cdot {\rm cm}^{-3}_{} \; , \\
\Omega & \approx & \Omega^{}_{K} / \sqrt{5} \; = \; 0.316 \times 10^{5}_{} \; M^{-1}_{} \; R^{-1.5}_{} ~ {\rm s}^{-1}_{}, \\
v & \approx & \alpha \Omega^{}_{K} r / \sqrt{5} \; = \; 0.933 \times 10^{10}_{} \; \alpha^{}_{} \; R^{}_{} ~ {\rm cm} \cdot {\rm s}^{-1}_{}.
\end{eqnarray}
Here $T$, $H$, $\rho$, $\Omega$ and $v$ are the temperature, half-thickness, density, angular velocity and radial velocity of the accretion flow respectively; $\alpha$ is the viscous parameter, $\dot{M}$ is the mass accretion rate in unit of $M^{}_{\sun} {\rm s}^{-1}$, $M$ is the black hole mass in unit of $M^{}_{\sun}$, and $R$ is the radius $r$ in unit of $r^{}_{s}$.

The accretion disk is made of $\alpha$-particles, neutrons, protons, electrons, positrons, photons, neutrinos and antineutrinos. The outer disk is mainly constituted of $\alpha$-particles, but once $T$ reaches about $10^{10}_{}$ K (happened at around $10^{2}_{} r^{}_{s}$), the photodisintegration breaks down $\alpha$-particles into neutrons and protons. The parameter $X^{}_{\rm nuc}$ is introduced to describe the mass fraction of free nucleons and then $1 - X{}_{\rm nuc}$ is the mass fraction of $\alpha$-particles. The mass fraction $X^{}_{\rm nuc}$ is found from the equation of nuclear statistical equilibrium \citep{2007ApJ...657..383C, 1994ARA&A..32..153M}
\begin{eqnarray}
4 \left [ Y^{}_{e} - \frac{\left ( 1 - X^{}_{\rm nuc} \right )}{2} \right ] \left [ 1 - Y^{}_{e} - \frac{\left ( 1 - X^{}_{\rm nuc} \right )}{2} \right ] \left ( 1 - X^{}_{\rm nuc} \right )^{-1/2}_{} \; = \; 1.55 \times 10^{-5}_{} ~ \rho^{-3/2}_{}  ~ T^{\; 9/4}_{} ~ \exp{\left ( - \frac{1.64 \times 10^{-9}_{}}{T} \right )} \; .
\end{eqnarray}
And the number densities of $\alpha$-particle, proton and neutron can be written as
\begin{eqnarray}
n^{}_{\alpha} & \approx &  \left ( 1 - X^{}_{\rm nuc} \right ) \frac{\rho}{4 m^{}_{p}} \; , \\
n^{}_{p} & \approx &  X^{}_{\rm nuc} Y^{}_{e} \frac{\rho}{m^{}_{p}} \; , \\
n^{}_{p} & \approx &  X^{}_{\rm nuc} \left ( 1 - Y^{}_{e} \right ) \frac{\rho}{m^{}_{p}} \; ,
\end{eqnarray}
where $Y^{}_{e}$ stands for the proton-to-baryon ratio
\begin{equation}
Y^{}_{e} \; = \; \frac{n^{}_{p}}{n^{}_{n} + n^{}_{p}} \; .
\end{equation}
The distributions of the electrons and the positrons are described by the Fermi-Dirac distribution
\begin{equation}
f^{}_{e^{\mp}_{}} (E^{}_{e}, \eta^{}_{e}) \; = \; \frac{1}{e^{E^{}_{e} / k^{}_{\rm B} T \mp \eta^{}_{e}}_{} + 1} \; .
\end{equation}
Here $\eta^{}_{e}$ is a dimensionless degeneracy parameter of electrons defined by $\eta^{}_{e} = \mu^{}_{e} / k^{}_{\rm B} T$, where $\mu^{}_{e}$ ($- \mu^{}_{e}$) is the chemical potential of electrons (positrons) and $k^{}_{\rm B} = 8.617 \times 10^{-11}_{} ~ {\rm MeV} \cdot {\rm K}^{-1}_{}$ is the Boltzmann constant. If $\eta^{}_{e}$ is much larger than unity, the electrons are strongly degenerate, whereas if $\eta^{}_{e} \ll 1$, the electrons are weakly degenerate and we can ignore the degeneracy. From Eq. (16) we can obtain the number densities of electrons and positrons
\begin{equation}
n^{}_{e^{\mp}_{}} \; = \; \frac{1}{\hbar^{3}_{} \pi^{2}_{}} \int^{\infty}_{0} \frac{1}{e^{\sqrt{p^{2}_{} c^{2}_{} + m^{2}_{e} c^{4}_{}} / k^{}_{\rm B} T \mp \eta^{}_{e}}_{} + 1} p^{2}_{} {\rm d} p \; .
\end{equation}
Since the disk matter is neutral, the charge neutrality requires
\begin{equation}
n^{}_{e^{-}_{}} - n^{}_{e^{+}_{}} \; = \; n^{}_{p} \; .
\end{equation}
Adopting the conclusion draw by \citet{2007ApJ...657..383C} that electrons and positrons are neither nondegenerate nor strongly degenerate at all radii, we simply choose $\eta^{}_{e} = 1$ at all radii and calculate
\begin{equation}
Y^{}_{e} = \left ( n^{}_{e^{-}_{}} - n^{}_{e^{+}_{}} \right ) \frac{m^{}_{p}}{\rho} \; .
\end{equation}
By taking this simplification, we avoid the iteration in our calculations. Note that at the outer disk the matter is dominated by $\alpha$-particles therefore $Y^{}_{e} = 0.5$ should be taken as a boundary condition. In our calculation we simply set $Y^{}_{e} = 0.5$ as the upper bound.

It has been found in previous works that for the hyperaccretion NDAF the most important heating and cooling processes are the viscosity, advection and neutrino cooling. The vertically integrated viscous heating rate (over a half-thickness $H$) is given by
\begin{equation}
Q^{+}_{} \; = \; Q^{+}_{vis} \; = \; \frac{3 }{8 \pi} ~ (\dot{M} \cdot M^{}_{\sun} {s^{-1}_{}}) ~ \Omega^{2}_{} \;.
\label{viscousheating}
\end{equation}
In the inner NDAF region, the advective energy transport rate $Q^{-}_{adv}$ can be approximately given by \citep{1994ApJ...428L..13N, 1995ApJ...438L..37A, 2002ApJ...579..706D}
\begin{equation}
Q^{-}_{adv} \; = \; \Sigma v T \frac{{\rm d} s}{{\rm d} r} \; \simeq \; \xi v T \frac{H}{r} \left ( \frac{11}{3} a^{}_{r} T^{3}_{} + \frac{4}{3} \cdot \frac{7}{8} a^{}_{r} T^{3}_{} + \frac{3}{2} \frac{\rho k^{}_{\rm B}}{m^{}_{p}} \frac{1 - X^{}_{\rm nuc}}{4} \right ) \; ,
\end{equation}
where $\Sigma = 2 \rho H$ is the surface density of the disk, $s$ denotes the specific entropy, $\xi$ is taken to be constant and equal to $1$ and $a^{}_{r} = 4 \sigma^{}_{\rm s} / c$ is the radiation constant. The three terms in the brackets are the entropy density of radiation, neutrinos, and gas, respectively. And in the outer ADAF region, $Q^{-}_{adv}$ can be approximately given by \citep{1999ApJ...516..420W}
\begin{equation}
Q^{-}_{adv} \; = \; - T \frac{\dot{M}}{4 \pi r} \frac{{\rm d} s}{{\rm d} r} \; \simeq \; \frac{\alpha}{\sqrt{5}} \; \rho \; H^{3}_{} \; \Omega^{3}_{K} \; \zeta \; ,
\end{equation}
where $\zeta = (4 - 0.75 \beta) \gamma^{}_{\rho} - (12 - 10.5 \beta) \gamma^{}_{T}$, $\beta$ is the ratio of gas to total pressure, $\gamma^{}_{\rho} = {\rm d} \ln \rho / {\rm d} \ln r$, and $\gamma^{}_{T} = {\rm d} \ln T / \ln r$. In this paper we take $\beta = 0$, since the radiation pressure dominates over the gas; thus $\zeta = 4 \gamma^{}_{\rho} - 12 \gamma^{}_{T} \simeq 1.5$.

In the NDAF, the neutrino or antineutrino production is dominated by the pair capture of electron and positron $p + e^{-}_{} \rightarrow n + \nu^{}_{e}$ \& $n + e^{+}_{} \rightarrow p + \bar{\nu}^{}_{e}$, which is more familiarly named as the URCA process \citep{2002ApJ...577..311K, 2002ApJ...579..706D, 2005ApJ...629..341K}. In the case that $E^{}_{e}$ is of the order of several MeV ($E^{}_{e} \ll m^{}_{n,p} c^{2}_{}$), the cross sections of the two processes of the URCA process are approximately the same
\begin{equation}
\sigma^{}_{e^{-}_{} p} \; \simeq \; \sigma^{}_{e^{+}_{} n} \; \simeq \; \frac{G^{2}_{\rm F} (1 + 3 g^{2}_{A})}{\pi^{}_{}} E^{2}_{\nu (\bar{\nu})} \; ,
\end{equation}
where $G^{}_{\rm F} = 2.302 \times 10^{-22}_{} ~ {\rm cm} \cdot {\rm MeV}^{-1}_{}$ is the Fermi coupling constant and $g^{}_{A} \approx 1.27$ is the nucleon axial weak charge. Here $E^{}_{\nu (\bar{\nu})}$ is the neutrino or antineutrino energy in the final state and we have $E^{}_{\nu} \approx E^{}_{e^{-}_{}} - Q$, $E^{}_{\bar{\nu}} \approx E^{}_{e^{+}_{}} + Q$ with $Q = (m^{}_{n} - m^{}_{p}) c^{2}_{} \approx 1.29$ MeV according to the energy conservation.
Taking into account the distributions in Eq. (16), we are able to write down the corresponding neutrino cooling rate
\begin{equation}
Q^{-}_{\nu} \; \simeq \;  \dot{q}^{}_{Ne} \cdot H \; = \; ( \dot{q}^{}_{e^{-}_{} p} + \dot{q}^{}_{e^{+}_{} n} ) \cdot H \; ,
\end{equation}
with
\begin{eqnarray}
\dot{q}^{}_{e^{-}_{} p} & = & \frac{G^{2}_{\rm F} (1 + 3 g^{2}_{A})}{2 \pi^{3}_{} \hbar^{3}_{} c^{2}_{}} \cdot n^{}_{p} \int^{\infty}_{Q} \frac{E^{}_{e} \sqrt{E^{2}_{e} - m^{2}_{e} c^{4}_{}} (E^{}_{e} - Q)^{3}_{}}{e^{E^{}_{e} / k^{}_{\rm B} T - \eta^{}_{e}}_{} + 1} \; {\rm d} E^{}_{e} \; , \\[3mm]
\dot{q}^{}_{e^{+}_{} n} & = & \frac{G^{2}_{\rm F} (1 + 3 g^{2}_{A})}{2 \pi^{3}_{} \hbar^{3}_{} c^{2}_{}} \cdot n^{}_{n} \int^{\infty}_{m^{}_{e} c^{2}_{}} \frac{E^{}_{e} \sqrt{E^{2}_{e} - m^{2}_{e} c^{4}_{}} (E^{}_{e} + Q)^{3}_{}}{e^{E^{}_{e} / k^{}_{\rm B} T + \eta^{}_{e}}_{} + 1} \; {\rm d} E^{}_{e} \; .
\end{eqnarray}
The emitted monochromatic neutrino (antineutrino) luminosity through the URCA process from a shell at $r$ with thickness ${\rm d} r$ and height $H(r)$ are given by
\begin{eqnarray}
{\rm d} L^{}_{\nu}(E^{}_{\nu}, r) & = & 2 \pi r {\rm d} r H(r) n^{}_{p} \frac{G^{2}_{\rm F} (1 + 3 g^{2}_{A})}{2 \pi^{3}_{} \hbar^{3}_{} c^{2}_{}} \frac{(E^{}_{\nu} + Q) \sqrt{(E^{}_{\nu} + Q)^{2}_{} - m^{2}_{e} c^{4}_{}} E^{3}_{\nu}}{e^{(E^{}_{\nu} + Q) / k^{}_{\rm B} T - \eta^{}_{e}}_{} + 1} \; {\rm d} E^{}_{\nu} \; , \\[2mm]
{\rm d} L^{}_{\bar{\nu}}(E^{}_{\bar{\nu}}, r) & = & 2 \pi r {\rm d} r H(r) n^{}_{n} \frac{G^{2}_{\rm F} (1 + 3 g^{2}_{A})}{2 \pi^{3}_{} \hbar^{3}_{} c^{2}_{}} \frac{(E^{}_{\bar{\nu}} - Q) \sqrt{(E^{}_{\bar{\nu}} - Q)^{2}_{} - m^{2}_{e} c^{4}_{}} E^{3}_{\bar{\nu}}}{e^{(E^{}_{\bar{\nu}} - Q) / k^{}_{\rm B} T + \eta^{}_{e}}_{} + 1} \; {\rm d} E^{}_{\bar{\nu}} \; .
\end{eqnarray}
And the total luminosity of neutrinos (antineutrinos) from the shell ${\rm d} r$ are given by
\begin{eqnarray}
L^{}_{\nu}(r) & = & \int^{\infty}_{0} \frac{{\rm d} L^{}_{\nu}(E^{}_{\nu}, r)}{{\rm d} E^{}_{\nu}} \; {\rm d} E^{}_{\nu} \; , \\ \nonumber\\
L^{}_{\bar{\nu}}(r) & = & \int^{\infty}_{Q + m^{}_{e} c^{2}_{}} \frac{{\rm d} L^{}_{\bar{\nu}}(E^{}_{\bar{\nu}}, r)}{{\rm d} E^{}_{\bar{\nu}}} \; {\rm d} E^{}_{\bar{\nu}} \; .
\end{eqnarray}

In the NDAF region, the energy loss by neutrino cooling dominates over the energy advection, therefore $Q^{+}_{vis} \approx Q^{-}_{\nu}$. While in the ADAF region, we have $Q^{+}_{vis} \approx Q^{-}_{adv}$, i.e., the viscous heating is balanced by the advection cooling.

\section{Global neutrino interaction}

When travelling inside the disk, the emitted neutrinos and antineutrinos may interact with the accretion flow matter and exchange energy and momentum with them. We firstly introduce the main interaction channels and the corresponding energy and momentum loss rate here \citep{2002astro.ph.11404B, 2002ApJ...577..311K, 2007ApJ...657..383C, 1996ApJS..102..411I}. Since the electron neutrinos (produced via the URCA process) are dominated over neutrinos of other two flavors, we consider in this section only the interactions of electron neutrinos/antineutrinos with the matter.

\begin{itemize}

\item neutrino absorption by nucleons: ~ $\nu^{}_{e} + n \rightarrow e^{-}_{} + p$ and $\bar{\nu}^{}_{e} + p \rightarrow e^{+}_{} + n$

This is the dominant interaction process in the NDAF region since there are plenty of nucleons there. If the energies of neutrinos and antineutrinos are relatively low ($\sim$ MeV) and $E^{}_{\nu} \ll m^{}_{p} c^{2}_{}$ is satisfied, the cross sections of above two processes are approximately the same:
\begin{equation}
\sigma^{}_{\nu n} (E^{}_{\nu}) \; \simeq \; \sigma^{}_{\bar{\nu} p} (E^{}_{\bar{\nu}}) \; \simeq \; \frac{G^{2}_{\rm F} (1 + 3 g^{2}_{A})}{\pi^{}_{}} E^{2}_{\nu (\bar{\nu})} \; \approx \; 9.85 \times 10^{-44}_{} E^{2}_{\nu (\bar{\nu})} ~ {\rm cm}^{2}_{} \; ,
\end{equation}
where $E^{}_{\nu (\bar{\nu})}$ is in unit of MeV. Note that, the reaction $\bar{\nu}^{}_{e} + p \rightarrow e^{+}_{} + n$ has a low threshold $E^{}_{\bar{\nu}} > 1.806 ~ {\rm MeV} \approx m^{}_{e} + m^{}_{n} - m^{}_{p}$. When neutrinos or antineutrinos are captured by the nucleons, all the energy and the momentum carried by neutrinos will be transferred to the flow matter.

\item neutrino-electron/positron elastic scattering: ~ $\nu + e^{-}_{}, e^{+}_{} \rightarrow \nu + e^{-}_{}, e^{+}_{}$

This is the dominant interaction process in the ADAF  region since there is little nucleons there. The processes of electron neutrinos scattering on electrons or positrons are proceeded via not only the neutral current interaction but also charged current interaction. The corresponding cross sections are given by \citep{2002astro.ph.11404B}
\begin{eqnarray}
\sigma^{}_{\nu e} (E^{}_{\nu}) & \simeq & \frac{3 G^{2}_{\rm F}}{2 \pi^{}_{}} \left ( 1 + 4 \sin^2 \theta^{}_{\rm W} + \frac{16}{3} \sin^4 \theta^{}_{\rm W} \right ) \left ( 1 + \frac{\eta^{}_{e}}{4} \right ) \frac{k^{}_{\rm B} T}{m^{}_{e} c^{2}_{}} E^{}_{\nu}\; \approx \; 5.57 \times 10^{-44}_{}  \left ( 1 + \frac{\eta^{}_{e}}{4} \right ) \frac{k^{}_{\rm B} T}{m^{}_{e} c^{2}_{}} E^{}_{\nu} ~ {\rm cm}^{2}_{} \; , \\[2mm]
\sigma^{}_{\bar{\nu} e} (E^{}_{\bar{\nu}}) & \simeq & \frac{3 G^{2}_{\rm F}}{2 \pi^{}_{}} \left ( \frac{1}{3} + \frac{4}{3} \sin^2 \theta^{}_{\rm W} + \frac{16}{3} \sin^4 \theta^{}_{\rm W} \right ) \left ( 1 + \frac{\eta^{}_{e}}{4} \right ) \frac{k^{}_{\rm B} T}{m^{}_{e} c^{2}_{}} E^{}_{\bar{\nu}} \; \approx \; 2.33 \times 10^{-44}_{}  \left ( 1 + \frac{\eta^{}_{e}}{4} \right ) \frac{k^{}_{\rm B} T}{m^{}_{e} c^{2}_{}} E^{}_{\bar{\nu}} ~ {\rm cm}^{2}_{} \; ,
\end{eqnarray}
where $\theta^{}_{\rm W}$ is the Weinberg angle and $\sin^2 \theta^{}_{\rm W} \approx 0.23$. The average energy transfer from neutrino to electron (positron) is found to be $(E^{}_{\nu(\bar{\nu})} - 4 k^{}_{\rm B} T) / 2$ by \cite{1975ApJ...201..467T}. We introduce the parameter $\theta$ to denote the average neutrino energy loss rate in neutrino-electron/positron scattering: $\theta \equiv (E^{in}_{\nu(\bar{\nu})} - E^{out}_{\nu(\bar{\nu})}) / E^{in}_{\nu(\bar{\nu})} \simeq 1/2 - 2 k^{}_{\rm B} T / E^{}_{\nu(\bar{\nu})}$. Since the typical energy of neutrinos  $E_{\nu(\bar{\nu})}$ is roughly equal to the typical energy of electron in the radius ($kT_e$) where neutrinos originate, this scattering always heats the disk matter, as in the NDAF region mentioned above. This is quite different from the case of global photons scattering in the case of optically thin ADAF \citep{2009ApJ...691...98Y}. In that case, the energy of photons can be larger or smaller than the average energy of electrons thus the scattering can play a heating or cooling role. As for the momentum transfer, the mean scattering angle of this process is quite large ($60^{\circ}_{} \sim 90^{\circ}_{}$), which means that even for relatively small energy transfer there can be a large momentum transfer in this scattering. In our numerical calculations, we simply choose the momentum transfer cross sections to be $\sigma^{tr}_{\nu e} = 2 \sigma^{}_{\nu e} / 3$.

\item neutrino-baryon elastic scattering: ~ $\nu + p, n, {\rm He} \rightarrow \nu + p, n, {\rm He}$

Neutrinos and antineutrinos interact elastically with nucleons ($n$, $p$) or the $\alpha$-particles via the neutral current interaction. The corresponding cross sections are the same for either neutrinos or antineutrinos and are approximately given by
\begin{eqnarray}
\sigma^{}_{p} (E^{}_{\nu(\bar{\nu})}) & \simeq & \frac{G^{2}_{\rm F}}{4 \pi^{}_{}} \left [ \left ( 1 - 4 \sin^2 \theta^{}_{\rm W} \right )^{2}_{} + 3 g^{2}_{A} \right ] E^{2}_{\nu(\bar{\nu})} \; \approx \; 2.04 \times 10^{-44}_{} E^{2}_{\nu(\bar{\nu})} ~ {\rm cm}^{2}_{} \; , \\
\sigma^{}_{n} (E^{}_{\nu(\bar{\nu})}) & \simeq & \frac{G^{2}_{\rm F}}{4 \pi^{}_{}} (1 + 3 g^{2}_{A}) E^{2}_{\nu(\bar{\nu})} \; \approx \; 2.46 \times 10^{-44}_{} E^{2}_{\nu(\bar{\nu})} ~ {\rm cm}^{2}_{} \; , \\
\sigma^{}_{\alpha} (E^{}_{\nu(\bar{\nu})}) & \simeq & \frac{16 G^{2}_{\rm F}}{\pi^{}_{}} \sin^4 \theta^{}_{\rm W} E^{2}_{\nu(\bar{\nu})} \; \approx \; 7.55 \times 10^{-46}_{} E^{2}_{\nu(\bar{\nu})} ~ {\rm cm}^{2}_{} \; .
\end{eqnarray}
Note that, the MeV neutrinos scatter from nuclei or nucleons almost coherently, which means neutrinos do not loss energy in these scattering. However, the mean scattering angles of these processes are large, which means there are also large momentum transfer in the scattering. In our calculations, we also simply choose the momentum transfer cross sections to be $\sigma^{tr}_{p, n, \alpha} = 2 \sigma^{}_{p, n, \alpha} / 3$ and ignore the energy transfer.
\end{itemize}

In order to calculate the ``global" neutrino heating rate at a given radius $r$, we need to know the received neutrino spectrum at $r$. The received neutrino (antineutrino) spectrum at $r$ emitted by the flow inside of $r$ is given by \citep{2009ApJ...691...98Y, 2007ApJ...655...88P}
\begin{equation}
F^{in}_{\nu (\bar{\nu})} (E^{}_{\nu (\bar{\nu})}, r) \; = \; \int^{r}_{r^{}_{in}} e^{- \tau^{}_{\nu (\bar{\nu})}}_{} \frac{1}{4 \pi r^{2}_{}} \frac{{\rm d} L^{}_{\nu (\bar{\nu})}(E^{}_{\nu (\bar{\nu})}, r')}{{\rm d} r'} {\rm d}r' \; ,
\end{equation}
where $r^{}_{in} = 3 r^{}_{s}$ is the inner edge of the NDAF, ${\rm d} L^{}_{\nu (\bar{\nu})}(E^{}_{\nu (\bar{\nu})}, r')$ has been given in Eqs. (27) and (28) and $\tau^{}_{\nu (\bar{\nu})}$ is the neutrino (antineutrino) optical depth from $r'$ to $r$
\begin{equation}
\tau^{}_{\nu (\bar{\nu})} \; = \; \int^{r}_{r'} (\sigma n)^{}_{\nu (\bar{\nu})} {\rm d}r'' \; ,
\end{equation}
with $(\sigma n)^{}_{\nu (\bar{\nu})}$ are functions of $r''$ and $r'$
\begin{eqnarray}
(\sigma n)^{}_{\nu} & = & \sigma^{}_{\nu n} (E'^{}_{\nu}) n''^{}_{n} + \sigma^{}_{p} (E'^{}_{\nu}) n''^{}_{p} + \sigma^{}_{p} (E'^{}_{\nu}) n''^{}_{n} + \sigma^{}_{\alpha} (E'^{}_{\nu}) n''^{}_{\alpha} + \sigma^{}_{\nu e} (E'^{}_{\nu}) (n''^{}_{e^{-}_{}} + n''^{}_{e^{-}_{}}) \; , \\[2mm]
(\sigma n)^{}_{\bar{\nu}} & = & \sigma^{}_{\bar{\nu} p} (E'^{}_{\bar{\nu}}) n''^{}_{p} + \sigma^{}_{p} (E'^{}_{\bar{\nu}}) n''^{}_{p} + \sigma^{}_{p} (E'^{}_{\bar{\nu}}) n''^{}_{n} + \sigma^{}_{\alpha} (E'^{}_{\bar{\nu}}) n''^{}_{\alpha} + \sigma^{}_{\bar{\nu} e} (E'^{}_{\bar{\nu}}) (n''^{}_{e^{-}_{}} + n''^{}_{e^{-}_{}}) \; ,
\end{eqnarray}
where $\sigma^{}_{\nu n, \; \bar{\nu} p, \; p, \; n, \; \alpha, \; \nu e, \; \bar{\nu} e}$ can be calculated by using Eqs. (31) - (36), in which $E'^{}_{\nu (\bar{\nu})}$ is the neutrino/antineutrino energy emitted at the radius $r'$ in unit of MeV, $T''$ is the temperature at  $r''$ and $n''^{}_{n, \; p, \; \alpha, \; e^{\mp}_{}}$ are the number densities of corresponding particles at the radius $r''$.
The neutrino (antineutrino) spectrum received at $r$ emitted by the flow outside of the rdius $r$ is given by \citep{2009ApJ...691...98Y, 2007ApJ...655...88P}
\begin{equation}
F^{out}_{\nu (\bar{\nu})} (E^{}_{\nu (\bar{\nu})}, r) \; = \; \int^{r^{}_{out}}_{r} \frac{e^{- \tau^{}_{\nu (\bar{\nu})}}_{}}{4 \pi r H(r')} \ln\sqrt{\frac{r' + r}{r' - r}} \frac{{\rm d} L^{}_{\nu (\bar{\nu})}(E^{}_{\nu (\bar{\nu})}, r')}{{\rm d} r'} {\rm d}r' \; .
\end{equation}
We set $r^{}_{out} = 10^4 r^{}_{s}$ in our calculations. The results are not sensitive to the exact value of $r^{}_{out}$. The total neutrino (antineutrino) spectrum received at $r$ is the sum of $F^{in}_{\nu (\bar{\nu})}$ and $F^{out}_{\nu (\bar{\nu})}$.
Then the global neutrino heating rate can be calculated by
\begin{eqnarray}
\dot{q}^{}_{global} & = & \int^{\infty}_{0} \left [ F^{in}_{\nu} (E^{}_{\nu}, r) + F^{out}_{\nu} (E^{}_{\nu}, r) \right ] \left ( \sigma^{}_{\nu n} (E^{}_{\nu}) n^{}_{n} E^{}_{\nu} + \sigma^{}_{\nu e} (E^{}_{\nu}) (n^{}_{e^{-}_{}} + n^{}_{e^{-}_{}}) \theta E^{}_{\nu} \right ) {\rm d} E^{}_{\nu} \nonumber \\
& & + \int^{\infty}_{Q + m^{}_{e} c^{2}_{}} \left [ F^{in}_{\bar{\nu}} (E^{}_{\bar{\nu}}, r) + F^{out}_{\bar{\nu}} (E^{}_{\bar{\nu}}, r) \right ] \left ( \sigma^{}_{\bar{\nu} p} (E^{}_{\bar{\nu}}) n^{}_{p} E^{}_{\bar{\nu}} + \sigma^{}_{\bar{\nu} e} (E^{}_{\bar{\nu}}) (n^{}_{e^{-}_{}} + n^{}_{e^{-}_{}}) \theta E^{}_{\bar{\nu}} \right ) {\rm d} E^{}_{\bar{\nu}} \; ,
\end{eqnarray}
where $n^{}_{n, \; p, \; e^{\mp}_{}}$ are the number densities at the radius $r$. The vertically integrated global neutrino heating rate is given by
\begin{equation}
Q^{+}_{global} \; = \; \dot{q}^{}_{global} \cdot H \; .
\label{globalheating}
\end{equation}

Besides the energy transfer from neutrinos to the flow matter, the emitted neutrinos/antineutrinos also transfer momenta to the flow matter when interacting with them, i.e., the radiated neutrinos exert pressure upon the flow matter. The average force suffered by the disk matter can be approximately evaluated by
\begin{eqnarray}
{\cal F}^{}_{\nu} (r) & = & \int^{\infty}_{0} {\rm d}E^{}_{\nu} \int^{r}_{r^{}_{in}} {\rm d}r' \; e^{- \tau^{}_{\nu}}_{} \frac{1}{4 \pi r^{2}_{} c} \frac{{\rm d}L^{}_{\nu}(E^{}_{\nu}, r')}{{\rm d} r'} \cdot \left [ \frac{n^{}_{p}}{n^{}_{tot}} \left (  \sigma^{tr}_{p}(E^{}_{\nu}) + \sigma^{tr}_{\nu e}(E^{}_{\nu}) \right ) + \frac{n^{}_{n}}{n^{}_{tot}} \left ( \sigma^{}_{\nu n} (E^{}_{\nu}) + \sigma^{tr}_{n}(E^{}_{\nu}) \right ) + \frac{n^{}_{\alpha}}{n^{}_{tot}} \left (  \sigma^{tr}_{\alpha}(E^{}_{\nu}) + 2 \sigma^{tr}_{\nu e}(E^{}_{\nu}) \right ) \right ] \nonumber \\[2mm]
& & + \int^{\infty}_{Q + m^{}_{e} c^{2}_{}} {\rm d}E^{}_{\bar{\nu}} \int^{r}_{r^{}_{in}} {\rm d}r' \; e^{- \tau^{}_{\bar{\nu}}}_{} \frac{1}{4 \pi r^{2}_{} c} \frac{L^{}_{\bar{\nu}}(E^{}_{\bar{\nu}}, r')}{{\rm d} r'} \cdot \left [ \frac{n^{}_{p}}{n^{}_{tot}} \left ( \sigma^{}_{\bar{\nu} p} (E^{}_{\bar{\nu}}) + \sigma^{tr}_{p}(E^{}_{\bar{\nu}}) + \sigma^{tr}_{\bar{\nu} e}(E^{}_{\bar{\nu}}) \right ) + \frac{n^{}_{n}}{n^{}_{tot}} \sigma^{tr}_{n}(E^{}_{\bar{\nu}}) + \frac{n^{}_{\alpha}}{n^{}_{tot}} \left ( \sigma^{tr}_{\alpha}(E^{}_{\bar{\nu}}) + 2 \sigma^{tr}_{\bar{\nu} e}(E^{}_{\bar{\nu}}) \right ) \right ] \; ,
\end{eqnarray}
where neutrinos are regarded to be massless and $n^{}_{tot} \equiv n^{}_{p} + n^{}_{n} + n^{}_{\alpha}$. This force is placed against the gravitational force of the central black hole
\begin{equation}
{\cal G}^{}_{p} (R) \; = \; G^{}_{N} \frac{(M \cdot M^{}_{\sun}) m^{}_{p}}{r^{2}_{}} \; ,
\end{equation}
where $G^{}_{N}$ is the gravitational constant.

\section{Numerical results }

We consider in this paper four accretion flow models: (a) $\dot{M} = 0.1 \; M^{}_{\sun} {\rm s}^{-1}_{}, \alpha = 0.1$; (b) $\dot{M} = 1.0 \; M^{}_{\sun} {\rm s}^{-1}_{}, \alpha = 0.1$; (c) $\dot{M} = 0.1 \; M^{}_{\sun} {\rm s}^{-1}_{}, \alpha = 0.01$; and (d) $\dot{M} = 1.0 \; M^{}_{\sun} {\rm s}^{-1}_{}, \alpha = 0.5$. We calculate the global neutrino heating rate numerically. Following the results in  \cite{2007ApJ...657..383C}, we set $r^{}_{\rm ign} = 60, 200, 450$ and $40 \; r^{}_{s}$ in Models (a), (b), (c), and (d), respectively. The four plots in Fig. 1 show the viscous heating rate $Q^{+}_{vis}$ (dashed), global neutrino heating rate $Q^{+}_{global}$ (solid), advection cooling rate $Q^{-}_{adv}$ (dotted), and neutrino cooling rate $Q^{-}_{\nu}$ (dot-dashed) for Models (a)-(d), respectively. Some quantities are not continuous across $r^{}_{\rm ign}$, this is because the analytical solutions of Eqs. (1)-(10) are not precise.

\begin{figure}
\begin{center}
\includegraphics[scale=0.75, angle=0, clip=0]{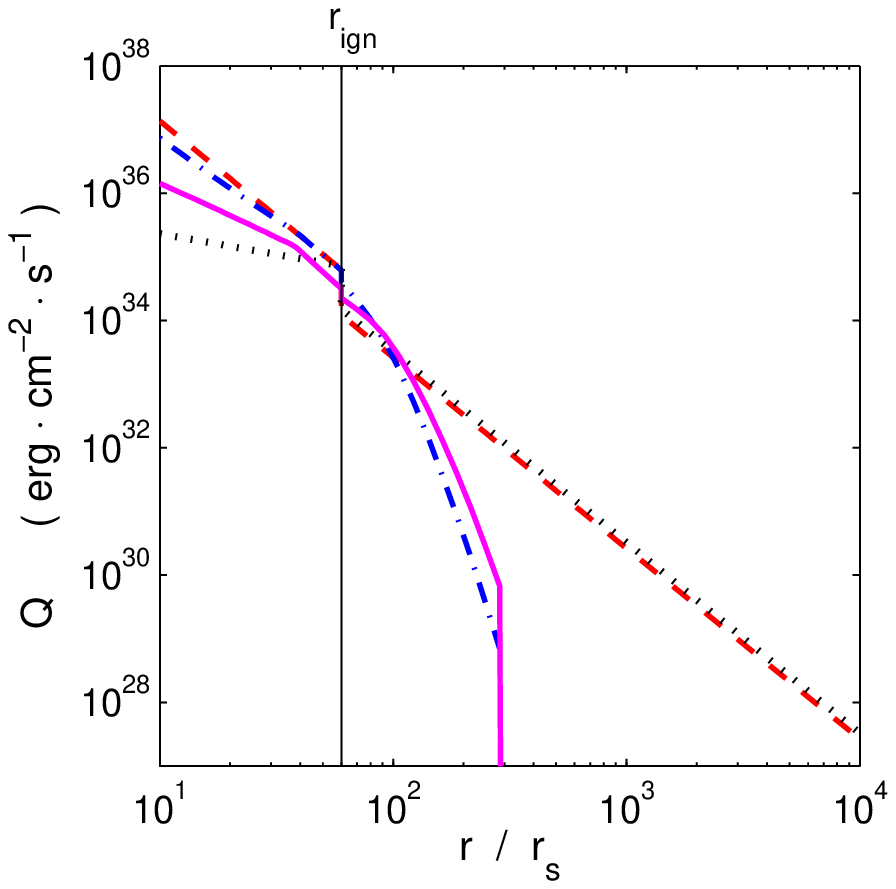}
\includegraphics[scale=0.75, angle=0, clip=0]{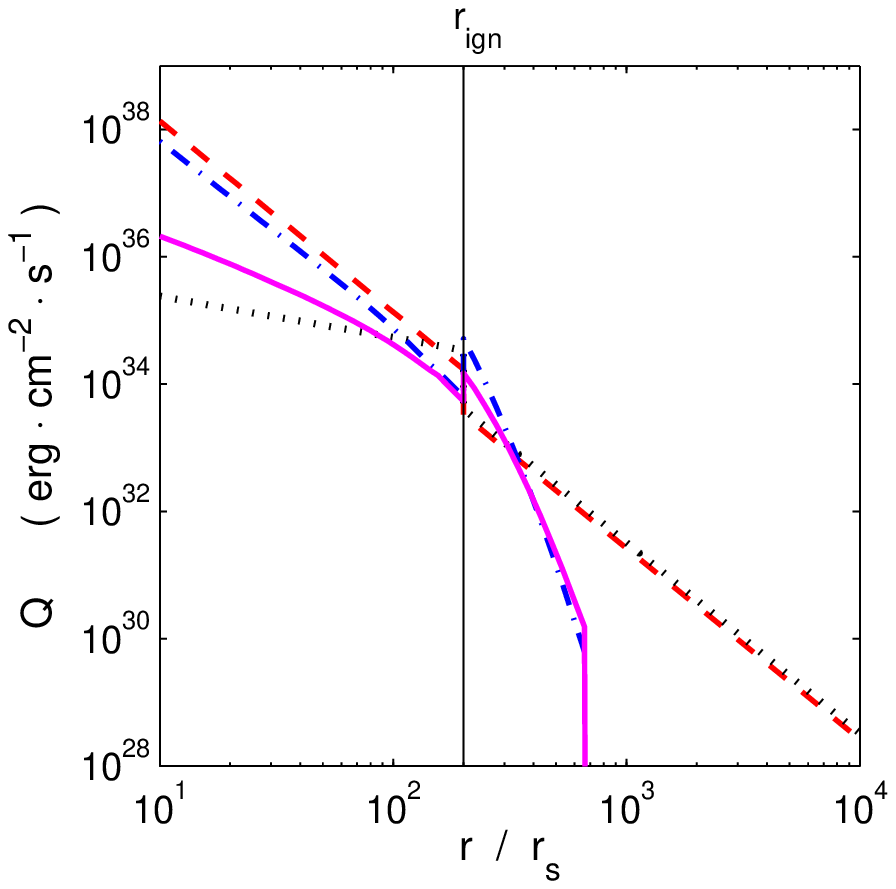}
\includegraphics[scale=0.75, angle=0, clip=0]{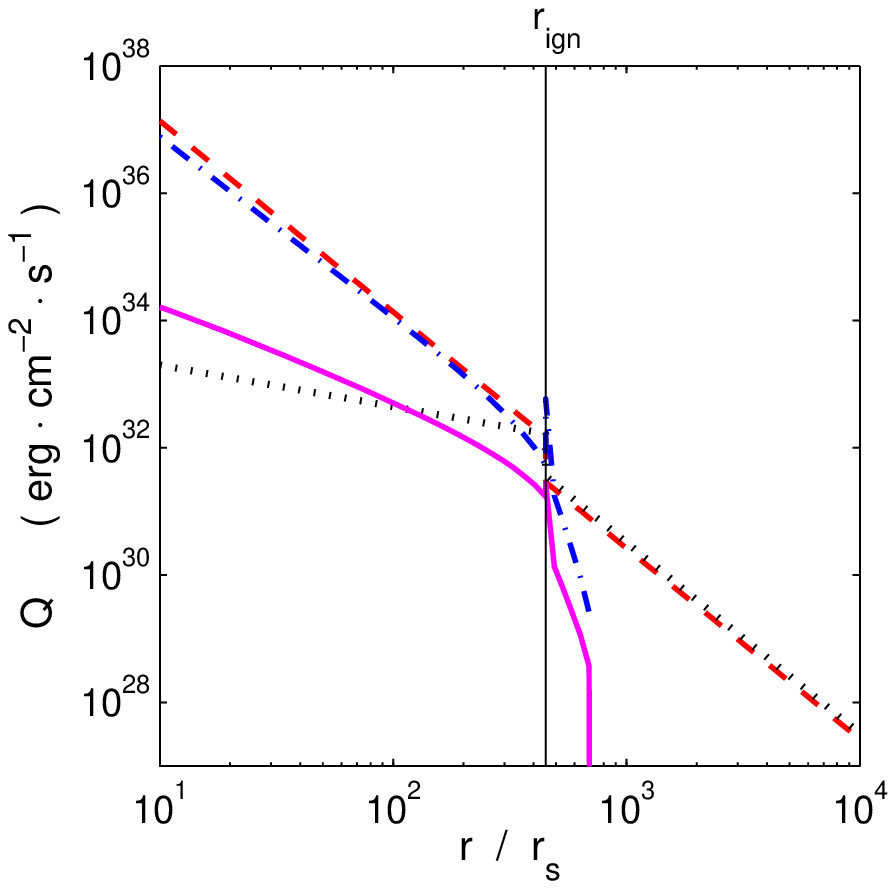}
\includegraphics[scale=0.73, angle=0, clip=0]{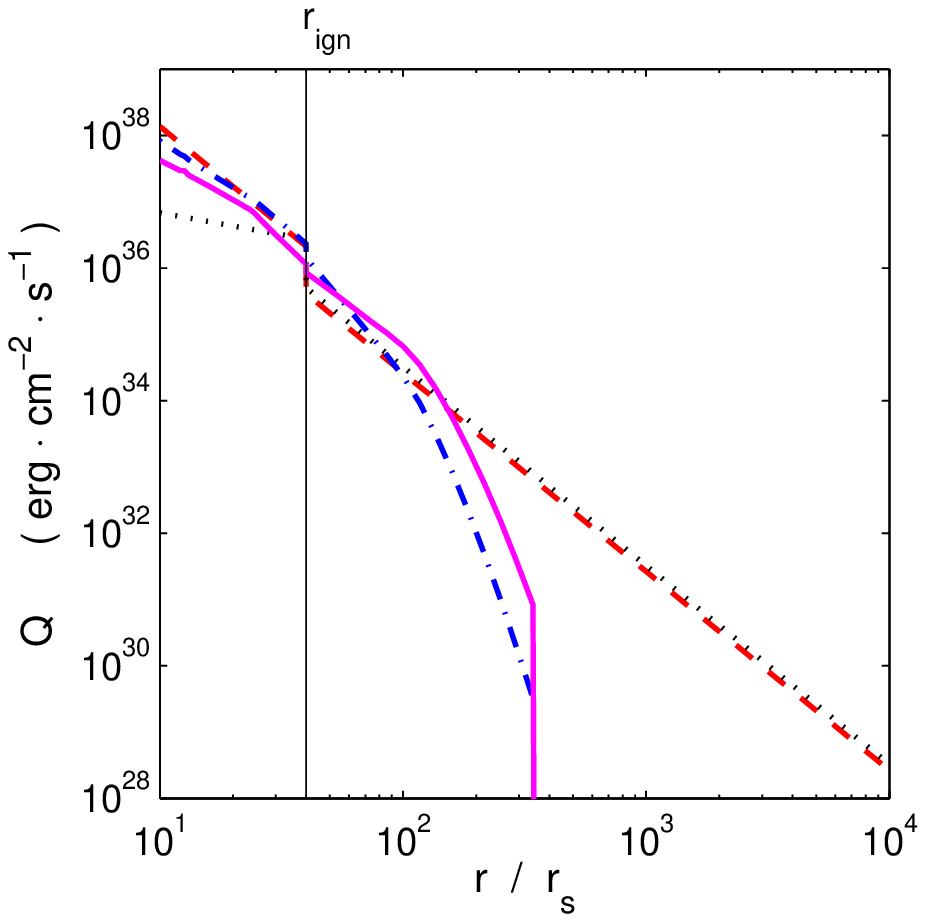}
\caption{The heating and cooling rates $Q^{+}_{vis}$ (dashed), $Q^{+}_{global}$ (solid), $Q^{-}_{adv}$ (dotted) and $Q^{-}_{\nu}$ (dot-dashed) as functions of radius $r$ for Model (a) ({\em top-left}; $\dot{M} = 0.1 ~ M^{}_{\sun} {\rm s}^{-1}_{}$, $\alpha = 0.1$, $r^{}_{\rm ign} \approx 60 \; r^{}_{s}$), Model (b) ({\em top-right}; $\dot{M} = 1.0 ~ M^{}_{\sun} {\rm s}^{-1}_{}$, $\alpha = 0.1$, $r^{}_{\rm ign} \approx 200 \; r^{}_{s}$), Model (c) ({\em bottom-left}; $\dot{M} = 0.1 ~ M^{}_{\sun} {\rm s}^{-1}_{}$, $\alpha = 0.01$, $r^{}_{\rm ign} \approx 450 \; r^{}_{s}$), and Model (d) ({\em bottom-right}; $\dot{M} = 1.0 ~ M^{}_{\sun} {\rm s}^{-1}_{}$, $\alpha = 0.5$, $r^{}_{\rm ign} \approx 40 \; r^{}_{s}$), respectively. }
\end{center}
\end{figure}

From Fig. 1 we can obtain the following results.
 \begin{itemize}

 \item  At the innermost region of the NDAF, the global neutrino heating rate is always much lower than the local viscous heating rate.

 \item However, with the increase of the radius, the ratio $Q^{+}_{global} / Q^{+}_{vis}$ increases, i.e., the global neutrino heating becomes more and more important. Physically, this is because the two terms have different scaling with radius. From Eq. (\ref{viscousheating}) we see that $Q^{+}_{vis}\propto r^{-3}$. The radial scaling of $Q^{+}_{global}$ is not easy to estimate. From Eq. (\ref{globalheating}) it is roughly proportional to $F^{}_{\nu (\bar{\nu})} n^{}_{n,p} r$. Here $F^{}_{\nu (\bar{\nu})}$ is the received neutrino flux at $r$, its radial scaling must be flatter than $r^{-2}$ because of the accumulation of neutrinos from all the radii within $r$. The number density $n^{}_{n,p}\propto r^{-2.55}$ from Eq. (\ref{density}). Thus it is quite possible to have the scaling of $Q^{+}_{global}$ flatter than $r^{-3}$.
 \item It is interesting to note that around the ``ignition'' radius $r^{}_{\rm ign}$, the global neutrino heating rate $Q^{+}_{global}$ becomes comparable to the viscous heating rate $Q^{+}_{vis}$ for Models (a), (b), and (d). For Model (d) the global heating even becomes larger than the viscous heating in the ADAF region. This result is especially of interest to us, as we will discuss below.
 \item When the radius is larger than $r^{}_{\rm ign}$, the global heating rate rapidly decreases with increasing radius. This is because the temperature decreases, thus $n^{}_{n,p}$ quickly decreases.
 \item By comparing the four plots in Fig. 1, we can see that the ratio $Q^{+}_{global} / Q^{+}_{vis}$ around $r^{}_{\rm ign}$ is more sensitive to $\alpha$ than to $\dot{M}$. A higher $\alpha$, rather than a higher $\dot{M}$, can more easily result in a larger ratio. Physically, this is because  the produced total neutrino flux in the NDAF is fully determined by $\dot{M}$. But a larger $\alpha$ implies a smaller density thus smaller radial optical depth, which then results in a stronger neutrino flux received at $r$.
 \end{itemize}

We can see that in some case such as Model (d), the global neutrino heating is very significant. It is  even larger than the viscous heating close to $r^{}_{\rm ign}$ in the ADAF region. In this paper we only focus on the evaluation of the significance of the global neutrino heating, but do not self-consistently calculate the dynamical structure and radiation of the hyperaccretion flow after the global neutrino effect is taken into account. However, two consequences can be expected: 1) the temperature of the disk at around $r^{}_{\rm ign}$ will be slightly raised and the ignition radius $r^{}_{\rm ign}$ will become larger. In this case the neutrino emission will be enhanced since the efficiency of the UCAR process is very sensitive to the temperature; 2) more importantly, if the global heating effect is important enough in the ADAF region, since the temperature of the flow in the ADAF is already close to the virial temperature, it is possible that the temperature of the ADAF will be higher than the virial one due to this global heating. In this case the accretion rate will be highly suppressed after the matter inside of $r^{}_{\rm ign}$ falls onto the black hole horizon. Only when the heated flow cools down or the unheated flow located at large radius comes in, does the accretion rate recover. This means that the black hole activity will oscillate between an active and an inactive phases. Such kind of behavior was first found by \citet{1978ApJ...226.1041C} and later was used to explain the intermittent activity of compact radio sources \citep{2011ApJ...737...23Y}. The duration of the inactive phases is determined by the cooling timescale of the heated flow, or the accretion timescale of the unheated flow located at large radius. The duration of the active phase is roughly determined by the accretion timescale of the NDAF at $r^{}_{\rm ign}$,
\begin{equation}
\tau^{}_{\rm act} \sim r / v \sim 5 \times 10^{-4} M^{1.2}_{} \alpha^{-1.2}_{} ( r^{}_{\rm ign} /r^{}_{s})^{0.8} \sim 1.1 ( M / 3 M^{}_{\sun} )^{1.2} ( \alpha / 0.1)^{-1.2} ( r^{}_{\rm ign} / 100 r^{}_{s})^{0.8}~{\rm s}
\end{equation}
Such a timescale is roughly equal to the ``slow'' variability timescale of the GRBs \citep{2012ApJ...748..134G}. If the duration of the inactive phase is comparable to or less than $\tau_{\rm act}$, the global neutrino heating mechanism may explain the origin of the ``slow'' variability component of GRBs; otherwise this may imply a challenge to the NDAF model.  A more exact global calculation with the global neutrino heating effect included is badly needed to obtain the self-consistent solution of the heated flow and thus make a final conclusion.

\begin{figure}
\begin{center}
\includegraphics[scale=0.7, angle=0, clip=0]{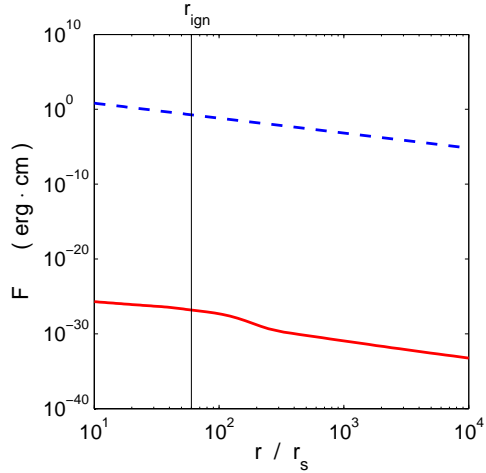}
\caption{The average neutrino radiation pressure ${\cal F}^{}_{\nu}$ (solid) on the disk matter with $\dot{M} = 0.1 ~ M^{}_{\sun} {\rm s}^{-1}_{}$ and $\alpha = 0.1$. The dashed line stands for the gravitation of the protons ${\cal G}^{}_{p}$ in the same accretion flow.}
\end{center}
\end{figure}

In addition to the energetics, the average neutrino radiation pressure on the disk matter is also calculated numerically. The results for Model (a) are illustrated in Fig. 2. One can see that the average neutrino radiation pressure ${\cal F}^{}_{\nu}$ (solid) is much lower than the gravitational force of the protons ${\cal G}^{}_{p}$ (dashed). We conclude that the neutrino radiation pressure in this hyperaccretion flow can be safely neglected.

\section{Summary and discussions}

In a NDAF model, most of the neutrinos are produced at the inner region of the accretion flow. Since the optical depth for neutrino is small in the radial direction, these neutrinos will be able to travel for a long distance and transfer their energy and momentum to the  accretion matter at large radii. This ``global neutrino interaction'' effect has been neglected in previous works and its significance is the focus of the present paper. Our numerical analysis shows that if $\alpha$ is not too small, $\alpha\ga 0.1$, the global neutrino heating rate will be comparable to or even larger than the local viscous heating rate at $r\ga r^{}_{\rm ign}$ (here $r^{}_{\rm ign}$ is the ``ignition'' radius, i.e., the boundary between the inner NDAF and the outer ADAF). Consequently, the temperature of the ADAF may become higher than the virial one. This implies that the flow outside of $r^{}_{\rm ign}$ can't be accreted and the accretion will be highly suppressed once the accretion flow inside of $r^{}_{\rm ign}$ is accreted. We found that the accretion timescale of NDAF at $r^{}_{\rm ign}$ is typically 1 second. This means that the NDAF model will not be able to explain the long GRBs. We would like to emphasize, however, that our calculations are based on the simple analytical solutions of NDAF and ADAF and are ``local'',  therefore the results should only be taken as preliminary. To get a more precise answer to the significance of the global neutrino interaction effect, a self-consistent and  global calculation with this effect included is required. The readers can refer to Yuan, Xie \& Ostriker (2009) for the details of such a calculation. Here we only briefly review the approach. The first step is to get the global solution of the set of differential equations describing the superaccretion flows (e.g., Popham et al. 1999). No global neutrino heating term is included in this step. Then we follow a similar approach as presented in this paper to calculate the global neutrino heating rate at each radius $\dot{q}^{}_{global}(r)$. Then we include this term in the energy equations of the accretion flow and try to solve their global solution again. Based on this new global solution we can calculate $\dot{q}^{}_{global}(r)$ again and compare with the result obtained in the last step. If they are not equal to each other, we replace $\dot{q}^{}_{global}(r)$ with the new one and repeat the above procedure until convergence is achieved.

In addition to the ``local'' nature of our calculation mentioned above, another caveat of our work is that we do not consider outflow in both NDAFs and ADAFs.  The radiation-megnetohydrodynamical numerical simulations to optically thick ADAFs show the existence of outflow \citep{2011ApJ...736....2O}. The mechanism is that the radiation force is found to exceed the gravitational force at the surface of the flow. We are not aware of any theoretical work focused on the study of possible outflow in NDAFs. However, we note that the ``micro Blandford-Payne'' mechanism recently proposed by \citet{2012ApJ...761..130Y} to explain the origin of outflow in an optically thin ADAF should work for any accretion flow in which the magnetic turbulence associated with the magnetorotational instability is the mechanism of transporting angular momentum. This mechanism is very similar to the Blandford \& Payne one in the sense that the magnetic centrifugal force drives the outflow. The difference is that no large-scale open field is required here (see also Bai \& Stone (2013, preprint) for the shearing box simulation analysis). The initial simulations to the radiatively efficient standard thin disk, which is physically similar to the NDAF, do show the existence of outflow \citep{2011ApJ...736....2O}. However, all these are our speculation since the parameters of accretion flows here are largely different from the works mentioned above. It is hard to estimate the effect of outflow on the significance of global neutrino interaction. On one hand, the inward decrease of accretion rate will make the neutrino production in the inner region of NDAF weaker in relative to the viscous heating rate at $r^{}_{\rm ign}$. This will make the global neutrino effect less important. But on the other hand,  the inward decrease of accretion rate means that the radial optical depth will become smaller which enhances the received neutrino flux at $r^{}_{\rm ign}$. The final result requires detailed numerical calculation.

\section{acknowledgments}

We thank Bing Zhang for valuable discussions and comments. S. L. and F. Y. are both supported by the National Basic Research Program of China (973 Program 2009CB824800). S. L. is supported by the National Natural Science Foundation of China (grant 11105113) and the Fujian Provincial Natural Science Foundation (grant 2011J05012). F. Y. is supported by the National Natural Science Foundation of China (grants 10821302, 11121062, and 11133005) and the CAS/SAFEA International Partnership Program for Creative Research Teams.

\end{document}